\documentclass[a4paper]{article}

\usepackage{INTERSPEECH2021}

\usepackage{todonotes}
\usepackage{acronym}
\newacro{VoxSRC-21}{Voxceleb Speaker Recognition Challenge 2021}
\newacro{SV}{speaker verification}
\newacro{DINO}{distillation with no labels}
\newacro{MoCo}{momentum contrast}
\newacro{SSL}{self-supervised learning}
\newacro{CV}{computer vision}
\newacro{AAM}{additive angular margin}

\usepackage{comment}
\usepackage{enumitem}
\usepackage{float}
\usepackage{graphicx}
\usepackage{multirow}

\title{The JHU submission to VoxSRC-21: Track 3}

\name{Jaejin Cho$^{1}$, Jes\'us Villalba$^{1,2}$, Najim Dehak$^{1,2}$}
\address{$^{1}$Center for Language and Speech Processing, $^{2}$Human Language Technology Center of Excellence, \\Johns Hopkins University, Baltimore, MD, USA}

\email{\{jcho52, jvillal7, ndehak3\}@jhu.edu}

\begin{document}

\maketitle
\begin{abstract}
This technical report describes Johns Hopkins University speaker recognition system submitted to Voxceleb Speaker Recognition Challenge 2021 Track 3: Self-supervised speaker verification (closed). Our overall training process is similar to the proposed one from the first place team in the last year's VoxSRC2020 challenge. The main difference is a recently proposed non-contrastive self-supervised method in \ac{CV}, \ac{DINO}, is used to train our initial model, which outperformed the last year's contrastive learning based on \ac{MoCo}. Also, this requires only a few iterations in the iterative clustering stage, where pseudo labels for supervised embedding learning are updated based on the clusters of the embeddings generated from a model that is continually fine-tuned over iterations.
In the final stage, Res2Net50 is trained on the final pseudo labels from the iterative clustering stage. This is our best submitted model to the challenge, showing 1.89, 6.50, and 6.89 in EER(\%) in voxceleb1\_test\_o, VoxSRC-21 validation, and test trials, respectively. %

\end{abstract}

\section{Introduction}
In~\ac{VoxSRC-21} Track 3: self-supervised~\ac{SV}, a participant is allowed to use only VoxCeleb2~\cite{Chung18voxceleb2} dev subset without any speaker labels for model training. For {SV} system validation, participants are restricted to use the VoxCeleb1~\cite{nagrani2017voxceleb1} pairs or the provided pairs composed of the subset of the VoxCeleb1 utterances where their distribution matches that of the test pairs. The validation pairs, however, cannot be used for training. The test pairs contain utterances that are shorter than the
utterances in training and validation subsets. 

This report shares the details about our developed systems and findings in this challenge. Our main focus was to check how the newly proposed non-contrastive \ac{SSL} method in~\ac{CV}, \ac{DINO}~\cite{caron2021dino}, works in speaker embedding learning. Thus, the training stages after that follow what's proposed from the first place team~\cite{thienpondt2020idlab} from the last year's challenge~\cite{nagrani2020voxsrc} with small modifications. This year's challenge has a special focus on multi-lingual verification but we do not develop specific systems for this.

\section{Method}
In short, the overall pipeline consists of training a front-end model for speaker embedding extractor and then scoring trial pairs with cosine scoring in the speaker embedding space. In the front-end modeling, we mainly have three stages similar to the first place team's development pipeline~\cite{thienpondt2020idlab} except using non-contrastive learning for the initial model. We explain the three stages in order.
\subsection{Initial model training: \ac{DINO}}
To generate pseudo labels from embedding clusters for supervised training, we train an initial model to extract good embeddings that can be clustered by speaker. We adopt a specific non-constrastive \ac{SSL} method, \ac{DINO}, for the purpose.
\subsubsection{Motivation: Why non-contrastive over constrative learning}
There have been many methods to train embeddings in a self-supervised manner~\cite{hsu2017unsupervised,stafylakis2019ss_spkemb,peng2020mixture,NEURIPS2020BYOL,huh2020augmentation,thienpondt2020idlab,xia2021moco_spkemb,caron2021dino}, and contrastive loss based methods with data augmentation are popular and well-performing ones~\cite{huh2020augmentation,thienpondt2020idlab,xia2021moco_spkemb}. The works using contrastive loss compose the negative samples with different samples from a current sample to make the current and negative samples far from each other in the embedding space. This, however, could be wrong as the size of the queue to accumulate negative samples gets bigger. For example, the more we accumulate utterances to compose negative samples, the more probable some of the utterances are from the same speaker of the current utterance. We could make the queue size smaller to avoid this issue but this degrades the performance~\cite{he2020mocov1}.
 In~\cite{xia2021moco_spkemb}, the author introduced a clustering stage at every epoch to sample negative samples only from different clusters but the improvement was not large.

Non-contrastive methods, however, do not require negative samples so they are free from this issue. Moreover, non-contrastive methods have shown comparable or even better performance compared to contrastive methods. Thus, we propose to apply a non-contrastive \ac{SSL} method recently proposed, DINO~\cite{caron2021dino}, that outperforms previous \ac{SSL} methods in many \ac{CV} tasks.

\begin{table*}[t]
\centering
\begin{tabular}{|c|c|c|c|c|c|}
\hline
\multirow{2}{*}{Stage}                    & \multirow{2}{*}{Algorithm/Loss} & \multirow{2}{*}{Model}             & \multicolumn{3}{l|}{EER   (\%)}         \\ \cline{4-6} 
                                          &                                 &                                    & voxceleb1\_o\_test & VoxSRC-21 val & VoxSRC-21 test \\ \hline
\multirow{2}{*}{\shortstack{Initial   model training \\ (self-supervised learning)}} & DINO                        & LResNet34                          & 4.83     & 13.96        &           -    \\ \cline{2-6} 
                                          & MoCo                       & ECAPA~\cite{thienpondt2020idlab}                              & 7.3     &       -       &      -        \\ \hline
\multirow{5}{*}{Iterative clustering}     & \multirow{5}{*}{AAM}         & ResNet34 (iter1)                   & 2.56     & 8.59         &         -      \\ \cline{3-6} 
                                          &                                 & ResNet34 (iter2)                   & 2.13     & 7.35         &        -       \\ \cline{3-6} 
                                          &                                 & ResNet34 (iter3) & 2.13     & 6.97         &         -      \\ \cline{3-6} 
                                          &                                 & ResNet34 (iter4)                   & 2.14     & 6.88         &        -       \\ \cline{3-6} 
                                          &                                 & ECAPA (iter7)~\cite{thienpondt2020idlab}                      & 2.1     &    -          &        -       \\ \hline
Robust training                           & \multirow{2}{*}{AAM}                          & \multirow{2}{*}{Res2Net50}                        & 1.89     & 6.50         & 6.88          \\ \cline{4-6}
+ larg-margin fine-tuning &                         &                          & 1.91     & 6.32         & 6.64*          \\ \hline
\end{tabular}
\caption{Speaker verification results over 3 different trial lists with progressing/different systems over the three stages. The numbers from~\cite{thienpondt2020idlab} seems rounded to the nearest tenth. Pseudo labels for robust training were generated from ResNet (iter3) model. * means the submission happened after the challenge deadline.}
\label{tab:all_results}
\end{table*}

\subsubsection{Distillation with No labels}
In~\cite{caron2021dino}, the author proposed a design to maximize the similarity between feature distributions of differently augmented images from an original image~. This is based on the assumption that augmented images from one image keep the same semantic information. For example, although you cropped two images from a dog image and make one a black image while making the other jittered, they are still dog images.

The training is done as follows: First, a given sample is augmented in different ways. To be specific, you crop a local view and a global view of an image, where local and global views mean small and large portions of the image. Several augmentations can be added to the cropped images, such as color jittering, Gaussian blur, solarization, etc. The local views are propagated through one branch while the global views are propagated through the other branch to minimize the cross-entropy between two distributions calculated along the branches. 
A student network in one branch and a teacher network in the other branch are initialized with the same architecture and the model parameters while they are updated in different ways during training. The student network is updated by gradient descent while the teacher network is updated by an exponential moving average of the student parameters. To avoid a model to find trivial solutions, i.e., having distributions where one dimension is dominant or uniform distributions, \textit{centering} and \textit{sharpening} are used. \textit{centering} prevents one dimension from dominating by calculating a center by equation. However, using \textit{centering} encourages a uniform distribution. That is why \textit{sharpening} is also applied where it encourages peaky distributions. This is done by setting a low value for the temperature in the teacher softmax normalization. The architecture for the student and teacher networks is composed of a backbone, e.g., ViT~\cite{dosovitskiy2021vit} or ResNet~\cite{he2016resnet} without later fully connected layers and a projection head. The projection head consists of a 3-layer fully connected layers with hidden dimension 2048 followed by L2 normalization and a weight normalized fully connected layer with $K$ dimensions.

\subsubsection{DINO to learn embedding from speech}
The assumption that augmented images from one image keep the same semantic information in \ac{CV} can be similarly applied to speaker embedding learning. For example, most of the speech corpora consist of utterances where each utterance is spoken by one speaker. In this case, it is reasonable to assume that segments extracted from random positions in the same utterance have the same speaker information. The correspondence between \ac{CV} and speaker embedding learning is following. An image corresponds to an utterance while cropping local and global views from an image corresponds to extracting short and long segments from an utterance. The popular augmentation methods in speaker embedding learning after extracting segments are adding sounds such as babbling, music, noise in the background, or applying room impulse response effects.

\subsection{Iterative clustering: pseudo label update}
In this stage, we train a new model based on pseudo labels generated from the initial model. In detail, we extract speaker embeddings from the initial model and cluster them using clustering algorithms where the number of clusters is heuristically determined based on speaker verification performance on validation data. Indices of final clusters are used as pseudo speaker labels for supervised speaker embedding training.

Once the first labels are generated from the initial model, we train a new model, possibly with a larger model. The model is continually updated over iterations based on pseudo labels updated after each iteration. The labels are updated in the same way explained above, i.e., through speaker embedding extraction, clustering, and pseudo labeling. The number of clusters is fixed as one value over the iterations.

\subsection{Robust training on the final pseudo-labels}
In this stage, a new larger model is trained with a large margin fine-tuning after a few epochs. The difference from the last year~\cite{thienpondt2020idlab} is that we keep using the \ac{AAM} loss instead of sub-center \ac{AAM}~\cite{deng2020sub} in this stage.

\section{Experiment and result}
For the input features in training, we used an 80-dimensional log filter bank calculated over the 25 ms window with a 10 ms shift. The moving window of 150 ms was used for the mean normalization of the features. Adam~\cite{kingma2014adam} optimizer with learning rate scheduling was used over the training stages.
\subsection{Initial model training: \ac{DINO}}
The embedding architecture we used for the DINO backbone is a  light version of ResNet34 (LResNet34) with the kernel size of the first convolution layer as 3 instead of 7 and with a mean and standard deviation pooling layer followed by a fully connected layer to have the embedding dimension as 256 as in~\cite{villalba2018jhumitNISTsre2018}. The $K$ in the following \ac{DINO} projection head was 65536. The reason for selecting the LResNet34 as the embedding architecture is to reduce the training time considering \ac{DINO} takes more computation compared to conventional supervised model training. For the augmentation, we added sounds such as babbling, music, noise in the background or applied room impulse response effects.

As shown the Intial model training (self-supervised learning) row in Table~\ref{tab:all_results}, \ac{DINO} outperforms \ac{MoCo} in the stage.

\subsection{Iterative clustering: pseudo label update}
In this stage, we used an original ResNet34 architecture~\cite{he2016resnet} with the kernel size of the first convolution layer as 3 instead of 7 and with a mean and standard deviation pooling layer followed by a fully connected layer to have the embedding dimension as 256. The loss function used was \ac{AAM} softmax~\cite{deng2019arcface}, warming up the margin value from 0 to 0.3 for the first 20 epochs. The number of clusters is set to 7500 following~\cite{thienpondt2020idlab}.

As shown in the Iterative clustering row in Table~\ref{tab:all_results}, the model performance converges from 2nd and 4th iterations on voxceleb1\_test\_o and VoxSRC-21 validation, respectively.
This is possibly because the embeddings from \ac{DINO} initial model are better than the ones from \ac{MoCo}.

\subsection{Robust training on the final pseudo-labels}
Res2Net50~\cite{gao2019res2net} architecture with 26 for the width of filters and 4 for the scale was used in the final robust training stage. The pooling layer and the following fully connected layers are the same in the previous stages. The \ac{AAM} loss with the same setting in the previous stage was used. The pseudo labels were generated from the 3rd model from the previous stage, ResNet34 (iter3) in Table~\ref{tab:all_results}. After 30 epochs, the post-pooling layers in the model were fine-tuned with a larger margin, 0.5. The large margin fine-tuned model on longer chunks, 4 seconds, however, degraded the performance on VoxSRC-21 test pairs although it showed improvement on validation pairs, as shown in Table~\ref{tab:lmft_durVSeer}. This is possibly due to the short-length segments, less than 4-second, that take a large portion of the utterances in the test trials. Thus, we used 3-second chunks instead for large margin fine-tuning.

\begin{table}[htpb]
\centering
\resizebox{\columnwidth}{!}{
\begin{tabular}{|c|c|c|c|}
\hline
Segment length & voxceleb1\_test\_o & VoxSRC-21\_val & VoxSRC-21\_test \\ \hline
No fine-tuning & \textbf{1.89}               & 6.50          & 6.88           \\ \hline
2 second       & 1.97               & 6.86          & -              \\ \hline
3 second       & 1.91               & 6.32          & \textbf{6.64}*           \\ \hline
4 second       & 1.92               & \textbf{6.31}          & 7.23           \\ \hline
\end{tabular}
}
\caption{Relationship between segment length and the performance as EER(\%) in large margin fine-tuning over 3 different trial lists. * means the submission happened after the challenge deadline.}
\label{tab:lmft_durVSeer}
\end{table}

Training a larger model and the following large margin fine-tuning improve the performance on VoxSRC-21 validation trials as shown in the Robust training row Table~\ref{tab:all_results}.

\section{Conclusion}
We developed speaker verification systems without using speaker labels and achieved 1.91, 6.32, and 6.64 in EER(\%) in voxceleb1\_test\_o, VoxSRC-21 validation and test trials, respectively. Our main difference from the previous year's first place team~\cite{thienpondt2020idlab} was to use non-contrastive self-supervised learning method, \ac{DINO}~\cite{caron2021dino}. This showed better performance in the initial model training stage compared to \ac{MoCo}~\cite{he2020mocov1}. Also, the better speaker embedding in the initial model led to only a few iterations in the next iterative clustering stage. In the robust training stage, we carefully chose the training segment lengths not to overfit to the training or validation subsets while avoiding too short segment lengths. This is because the large portion of the utterances in the test trials was shorter than ones in the training and validation pairs, having the lengths less than 4 seconds.

\vspace{-2mm}
\newpage
\bibliographystyle{IEEEtran}
\bibliography{main}

\end{document}